\documentclass{PoS}

\usepackage{
  graphicx,graphics,psfrag,amsbsy,amsmath}

\newcommand{\noi}{\noindent} 

\title{High density effective theory on the lattice}
\ShortTitle{High density effective theory on the lattice}

\author{A. Dougall\\
        Bergische Universit\"{a}t Wuppertal, 42119 Wuppertal, Germany\\
        E-mail: \email{ajd1@theorie.physik.uni-wuppertal.de}}

\abstract{Long-range interactions in finite density QCD necessitate a
non-perturbative approach in order to reliably map out the key features
and spectrum of the QCD phase diagram. However, the complex nature of 
the fermion determinant in this sector prohibits the use of established 
Monte Carlo techniques that utilize importance sampling. Whilst significant 
progress has been made in the low density, high temperature region,
this remains a considerable challenge at mid to high density. At large
chemical potential, QCD can be approximated using high density effective 
theory which is free from the sign problem at leading order. We investigate 
the implementation of this theory on the lattice in conjunction with
existing re-weighting techniques.}

\FullConference{The XXV International Symposium on Lattice Field Theory\\
                 July 30 - August 4 2007\\
                 Regensburg, Germany}

\begin{document}

\section{Introduction}
A key motivation for studying QCD at different temperatures and pressures 
arises from the need to understand the behaviour of matter in various
extreme scenarios: the environment of intense heat and pressure that existed 
after the big bang, the ongoing experiments in heavy ion collisions, such as 
RHIC and ALICE, that aim to recreate a similar environment within the 
laboratory, and the conditions that exist at the centre of a neutron star 
where matter is cold and very densely packed. Furthermore, by investigating 
the different phases of QCD, one may gain further insight into confinement, 
chiral symmetry breaking and the nature of the QCD vacuum, which would 
contribute to our understanding of the structure of hadrons. 

\begin{center}
\begin{figure}[hhh]
  \psfrag{neutron stars}{neutron stars}
  \psfrag{n}{nuclear}
  \psfrag{m}{matter}
  \psfrag{h}{hadron}
  \psfrag{g}{gas}
  \psfrag{v}{vacuum}
  \psfrag{cf1}{CFL}
  \psfrag{2sc}{2SC}
  \psfrag{t}{$T$}
  \psfrag{tc}{$T_c$}
  \psfrag{mu}{$\mu$}
  \psfrag{chiralon}{$\langle \bar{\psi}\psi \rangle > 0$}
  \psfrag{chiraloff}{$\langle \bar{\psi}\psi \rangle \sim 0$}
  \psfrag{q}{QGP}
  \hspace{1.5cm}
  \includegraphics[scale=0.55]{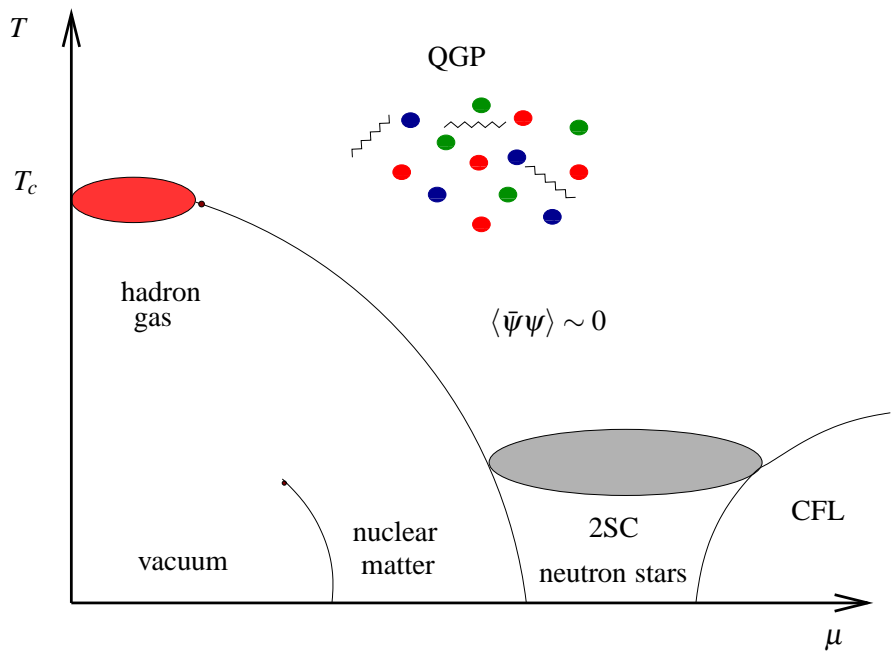}
  \caption{The QCD phase diagram.}
  \label{phasediag}
\end{figure}
\end{center}

The behaviour of strongly interacting matter in thermal and chemical 
equilibrium is characterised by the temperature $T$ and quark chemical 
potential $\mu$. The main thermodynamic properties of QCD are summarised 
in the QCD phase diagram, shown in fig.~\ref{phasediag}.

\vspace{-0.5cm}
Early attempts to map out the phase diagram revealed the existence of a 
deconfined phase and since then further details about the nature and location 
of the transition that separates these two phases have been uncovered. Some 
key features of the diagram include: 

\begin{itemize}
\item confined hadronic matter and the QCD phase transition: 
  We live in the phase where chiral symmetry is broken and there is colour 
  confinement. At low temperatures, there is a boundary separating 
  nuclear matter and the vacuum. As the temperature increases, 
  a hadronic gas forms. If the temperature or pressure increases beyond 
  certain limits, the system undergoes a phase transition, 
  the nature of which depends on whether this is driven by an 
  increase in temperature or density, or, an increase in both. 
\item Quark-Gluon Plasma: 
  In this phase QCD is deconfined and quarks and gluons are 
  the fundamental degrees of freedom. In addition, chiral symmetry is 
  approximately restored.

\item temperature-driven phase transition: 
  Located alongside the temperature axis of the phase diagram, the 
  transition here was recently found to be a crossover 
  \cite{Aoki:2006we}, 
  which places contraints on evolutionary models of the early universe.

\item critical temperature: Located at the point where the 
  phase transition crosses the temperature axis, the value of the critical 
  temperature was recently calculated \cite{Aoki:2006br} to 
  be $T=150(3)(3)$MeV although since the phase transition is 
  non-singular, there is no unique value for this quantity.

\item density-driven phase transition: 
  This transition, located along the chemical potential axis, is expected 
  to be first order, however, it is currently difficult to reach due to its 
  location in a region of intermediate density.

\item critical points: A critical point is expected to occur 
  along the first order phase transition as it moves towards a 
  temperature-driven crossover. Lattice predictions \cite{Fodor:2004nz}
  locate 
  this point at $(T,\mu)$ = (162(2),360(40)) MeV (for $N_t$ = 4 and with 
  large cut-off corrections expected). Recent work \cite{Fodor:2007vv} 
  indicates the existence of a 
  triple-point connecting three different phases on the diagram,
  occuring at $(T,\mu_q)_\mathrm{tri}\approx (137,300)$MeV (with large cut-off 
  corrections). 
\end{itemize}

\noi
In addition to the QGP phase, analytic calculations in the cold dense region 
of QCD have found alternative interaction channels leading to the prediction 
of colour superconductivity (2CS), where an attraction between pairs of quarks 
at the Fermi surface with opposite momenta leads to the formation of a 
condensate of quark cooper pairs. More recently, the phenomena 
of colour-flavour locking (CFL) was predicted \cite{Alford:1998mk}, where 
diquark condensates form from three light quark flavours, breaking colour and 
flavour symmetries. Given that these interactions are highly 
non-perturbative, one would like to investigate these regions, together 
with the density-driven phase transition using lattice QCD. However, the 
sign problem has proven to be a considerable barrier to the investigation 
of densely packed matter. 

\section{The sign problem}
In the lattice formulation, the expectation value of the operator 
$O$ is expressed in terms of the discrete Path Integral (PI). 
Having integrated over the fermion fields, the expectation value is given
by
\begin{eqnarray*}
  \langle O \rangle
  = \frac{\int D U 
    O\, \left(\det M \right)^{n_f}\,e^{-S_{G}}}   
  {\int D U \, 
    \left(\det M 
    \right)^{n_f}\,e^{\,-S_{G}}} 
\end{eqnarray*}
where $M$ is the fermion matrix and $n_f$ is the number of fermion flavours. 
This integral is computed numerically using standard Monte Carlo integration 
and in order to ensure that the background field configurations are 
sampled as efficiently as possible, importance sampling techniques are 
introduced, where the probability of selecting a particular configuration 
is weighted according to the distribution 
$P[U] = \frac{1}{Z}\left(\det M \right)^{n_f}\, e^{-S_{G}}$. The PI thus 
reduces to a sum over the operator evaluated on each background 
configuration, but we now have the constraint that $\det M$ be interpreted as 
a probability and therefore the Dirac operator must be positive. We can 
establish whether this is indeed the case by testing the 
``$\gamma_5$-hermiticity'' property: $M^\dagger = \gamma_5 M \gamma_5$.
If this equation is satisfied, either $\det M$ is real or the eigenvalues 
are complex but paired, and the fermion determinant can be used to importance 
sample the background gauge configurations. In the case of QCD at non-zero 
density, the $\gamma_5$-hermiticity is spoiled by the sign of the chemical 
potential term and importance sampling is prohibited. Whilst progress has 
been made in the low density region, this remains a problem at high density 
and we are motivated to approach the problem using an effective theory.

\section{Effective theory approach to QCD at high density}
In this section we examine the physical characteristics of densely packed 
matter and introduce the high density effective theory (HDET) that 
incorporates the low-energy behaviour of the system. This theory was 
introduced and developed by Hong in a series of papers, the first of which is 
given in Ref. \cite{Hong:1998tn}. 

\subsection{The physical characteristics of densely packed matter}
\begin{center}
  \begin{figure}[hhh]
    \hspace{3cm}
    \psfrag{a}{vacuum}
    \psfrag{d}{Dirac sea}
    \psfrag{f}{Fermi sea}
    \psfrag{mu}{$\mu$}
    \psfrag{E}{$E>\mu$}
    \includegraphics[scale=0.36]{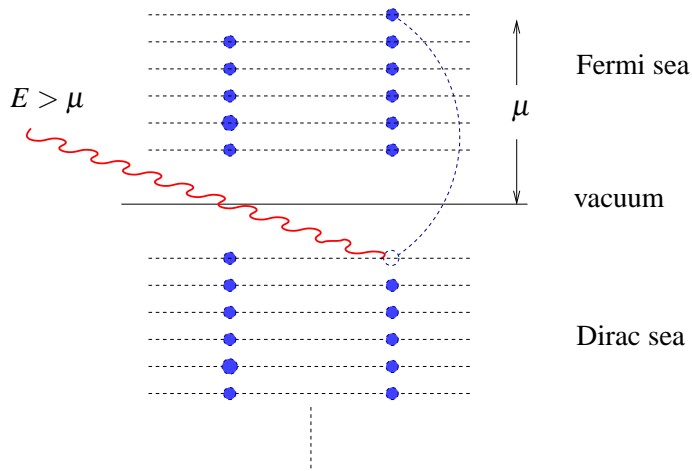}
    \caption{States within the Dirac sea and deep within the Fermi sea 
      do not contribute to the low energy dynamics due to Pauli blocking.}
    \label{pauli}
  \end{figure}
\end{center}

When matter is added to a fixed volume the energy of the lowest available 
state increases due to the fermionic nature of quarks. Consequently, quarks 
at the Fermi surface of very densely packed matter have high momentum. An 
essential characteristic of such a system is that at low energy, high-momentum
gluon exchange between quarks is suppressed because of 
the asymptotic nature of the QCD coupling. Typical Fermi surface 
interactions only change the quark momentum by a small amount and since these 
interactions are soft, $O(\Lambda_{\mathrm{QCD}})$, they require a 
non-perturbative treatment. Furthermore, the low energy degrees of 
freedom are restricted since quarks at low momentum (within the Fermi 
sea) and antiquarks (within the Dirac sea) require high momentum interactions 
to excite them due to Pauli blocking (occupation) of the states above 
(see fig.~\ref{pauli}). In essence, the only low energy dynamical degrees of 
freedom are the modes that lie near the Fermi surface which are quasi quarks 
and holes, together with soft gluons.
\vspace{-1cm}

\subsection{Approximating dense QCD with an effective theory}
Analogously to HQET, since there is a high momentum scale which does not 
contribute to the low energy dynamics, in this case near the Fermi surface, 
the quark momentum can be decomposed into two pieces, 
$p^\mu = \mu v_F^\mu + l^\mu$, where $v_F$ is the Fermi velocity and the 
residual part $l^\mu$ is associated with the Fermi surface interactions. 
In order to access these modes, the full field is expanded into large and 
small components (quasi-quarks) to give
\begin{eqnarray}
  \Psi(x) = 
  \sum_{\vec{v_F}}
  e^{-i\mu v_F \cdot x}\left(
  \psi_+(\vec{v_F},x) + \psi_-(\vec{v_F},x) \right)
  \label{decom}
\end{eqnarray}
where the large contribution to the momentum has been factored out
leaving field components $\psi_+$ and $\psi_-$ that are velocity
dependent and carry residual momentum $l$.  The large and small components
are defined respectively as
\begin{center}
  $\psi_+(x)=e^{i\mu v_F \cdot x} P_+ \Psi(x)$\,\, and \,\,
  $\psi_-(x)=e^{i\mu v_F \cdot x} P_- \Psi(x)$  
\end{center}
with the projection operator 
$P_\pm=\frac{1}{2}(1 \pm \vec{\alpha}\cdot \hat{v})$ and
$\vec{\alpha}=\gamma^0 \vec{\gamma}$. Since different modes of the quark 
field are characterised by an associated velocity, quarks with 
momentum modes that differ by more than the low energy exchanges 
decouple from each other and the Fermi surface (sphere) describing 
all the possible momentum modes can be split into patches, each 
labelled by an associated velocity. The size of a patch must be 
large enough to contain the low energy interactions which 
are $O(\Lambda_{QCD})$. It is possible to define the decomposition of the 
quark field into Fermi surface modes more precisely \cite{Hong:2002nn}, 
thus avoiding the sum over patches, however, we do not use this approach 
since the resulting action cannot be discretised readily in its present form. 

Using eqn.~\ref{decom} the full QCD Lagrangian at high density can be 
re-expressed in terms of large and small components, with the $\psi_+$ 
field describing quark modes at the Fermi surface and the $\psi_-$ field
describing modes within the Dirac sea. Soft interactions cannot excite 
$\psi_-$ so these modes are integrated out using the equations of motion 
to form a low energy effective theory that describes the strong interactions 
of modes at the Fermi surface. The tree-level effective Lagrangian of high 
density QCD \cite{Hong:1998tn} is given by
\begin{eqnarray}
L^0_{\mathrm{hdet}} =
    \sum_{\bf v} 
    \overline{\psi}_{\bf v} \left[
      i \gamma_\parallel^\mu D_\mu
      -
      \frac{\gamma_0 (D_\perp)^2}{2\mu}
      \sum_{n} \left( -\frac{iD_\parallel}{2\mu} \right)^n 
      \right]\psi_{\bf v}
    -\frac{1}{4}F_{\mu\nu}^aF^{a \,\mu\nu} 
\end{eqnarray}
where the parallel and perpendicular components of a generic four-vector 
$X^\mu$ are denoted $X_\parallel^\mu = V^\mu X\cdot V$ and 
$X_\perp^\mu = X^\mu -X_\parallel^\mu$ respectively, with 
$V = (1,\vec{v_F})$. The 
fields are now labelled by ${\bf v}$ instead of $+$, reflecting the
fact that they are velocity dependent and the index ${\bf v}$ runs over 
the number of patches. The HDET Lagrangian is a systematic 
expansion in $1/\mu$ and the coupling constant $\alpha_s$. There 
are coefficients in front of each term which contain information about the 
short distance physics which has been integrated out. These corrections are 
computed by matching the diagrams between the effective theory 
and the full theory at a given order in $\alpha_s$.  The leading order term 
in the fermion action can be re-expressed as $ iV\cdot D $ and is therefore 
spin-independent, whilst the NLO term can be written as a sum of 
spin-independent and spin-dependent terms.

The full dense QCD Lagrangian has now been re-expressed as an effective
Lagrangian with $\mu$ appearing in the denominator. Provided that the 
residual momentum is much smaller than the chemical potential, this should
serve as a good approximation to QCD at high density. Crucially, the sign
problem does not occur at leading order in this theory \cite{Hong:2002nn}, 
thus opening up the possibility for lattice calculations at high density.

\section{HDET on the lattice}
The strategy of the lattice calculation is to use the effective theory to
look for evidence of $\mu$-dependence in a simple observable such as the 
plaquette. In formulating HDET on the lattice we highlight several important 
details concerning the inclusion of the NLO term, an approximation 
to the sum over velocities and the discretisation of the action.

\subsection{Reweighting the leading order calculation}
As discussed, the LO fermion operator is free from the sign problem and
can therefore be used in the importance sampling of the field configurations. 
This is not possible in the case of the NLO term where the sign problem
re-emerges. Since it is necessary to include this correction in order to 
have explicit $\mu$-dependence in the calculation, as well as enabling 
measurements at lower values of $\mu$, this term is incorporated into the 
measurement by reweighting the observable \cite{Fodor:2001pe}, such that
\begin{eqnarray*}
  \langle O  \rangle = 
  \frac{\langle O \,e^{-S_\mathrm{NLO}}\rangle}{
  \langle e^{-S_\mathrm{NLO}}\rangle} .
\end{eqnarray*}

\subsection{The sum over patches}
The effective Lagrangian contains a sum over patches, each of which correspond
to a particular region of the Fermi surface with an associated unit velocity, 
$\vec{v}$. The physics within each patch is equivalent and therefore, to avoid a prohibitively expensive and complicated computation, we compute the 
Lagrangian for a single patch and then multiply by an overall factor to 
account for the sum. The number of patches covering the surface is given 
approximately by $N_{patch} \sim(4\pi\mu^2/\Lambda_\perp^2)$ where 
$\Lambda_\perp$ 
is the cutoff on the transverse momenta. For example, given a chemical 
potential of 1.5GeV, the number of patches is approximately 315.
The fermion part of the HDET Lagrangian (through NLO) becomes 
\begin{eqnarray*}
  L^0_{\mathrm{hdet}} =
  N_\mathrm{patch}
  \overline{\psi}_{\bf v} \left[
    i \gamma_\parallel^\mu D_\mu
    -
    \gamma_0\frac{(D_\perp)^2}{2\mu}
      \right]\psi_{\bf v}
\end{eqnarray*}
and we choose $\vec{v} = (0,0,1)$. 

\subsection{Discretisation}
The discretisation of the HDET fermion action follows from the usual definition
for the simplest form of the covariant derivative on the lattice. 
For $\vec{v} = (0,0,1)$, the LO term is given by
\vspace{-0.05cm}
\begin{eqnarray}
  N_\mathrm{patch} V \cdot D =  \frac{N_\mathrm{patch}}{2}\sum_{\mu=z,t} \left[ U_\mu(x) \delta_{x,y-\mu}
    - U^\dagger_\mu(x-\mu)\delta_{x,y+\mu}  \right]
\end{eqnarray}
where $V = (1,\vec{v})$. The NLO term can be written as
\begin{eqnarray}
  M_\mathrm{NLO} &=& 
  \frac{ N_\mathrm{patch}}{2\mu}
  \left( D_x^2 + D_y^2 + i\sigma_z [D_x,D_y] \right)
\end{eqnarray}
where $\sigma_z$ is the usual Pauli matrix.
\section{Discussion}
The LO operator that is used to generate the configurations has been 
tested for reversibility and linearity of $\Delta H$ with the 
microcanonical step size squared. Reweighting in the NLO term has also been 
incorporated and measurements are under way. The HDET code has 
been adapted from the MILC code \cite{milc}. One remaining issue in the 
implementation of HDET on the lattice is the inclusion of the Debye 
screening mass term which should be included in order to match to the full
theory \cite{Hong:2004qf}. Since this contribution is not gauge invariant, 
it is not clear how this term can be included in the generation of dynamical
lattices and more work must be carried out in this area.

In summary, we have given a brief overview of the QCD phase diagram 
and highlighted the necessity of finding an alternative approach to QCD
at mid to high density on the lattice. We have reviewed the principle 
features of high density effective theory and presented our strategy for
implementing this theory on the lattice. The aim of this work is to look 
for evidence of the density-driven phase transition. It is not clear at this 
stage how low it is possible to take the chemical potential before the 
breakdown of the expansion in the effective theory, however, we are guided 
by the fact that provided the residual momentum $l$ which is 
$O(\Lambda_\mathrm{QCD})$ is much smaller than the chemical potential, 
ie. $l/\mu<<1$, the expansion should be valid. 
\acknowledgments{
I am very grateful to Zoltan Fodor, who proposed this project,
Deog Ki Hong, Craig McNeile, Kalman Szabo and Fermin Viniegra for many 
useful discussions.}

\end{document}